\documentclass[aps,prl,twocolumn,showpacs,superscriptaddress,amsmath,amssymb,amsfonts,floatfix]{revtex4}

\usepackage[pdftex]{graphicx}
\usepackage{color}
\usepackage{textcomp}

\renewcommand{\section}[1]{\textit{#1} --}

\newcommand{\avg}[1]{\langle #1 \rangle}
\newcommand{\abs}[1]{\vert #1 \vert}

\newcommand{\Cb}{\mathbf C}
\newcommand{\Css}{C^\infty}
\newcommand{\Zb}{\mathbf Z}
\newcommand{\Yb}{\mathbf Y}
\newcommand{\Pb}{\mathbf P}
\newcommand{\Kb}{\mathbf K}
\newcommand{\bz}{\mathbf 0}
\newcommand{\bo}{\mathbf 1}

\begin{document}

\title{Scaling Theory of Heat Transport in Quasi-1D Disordered Harmonic Chains}

\author{Joshua D. Bodyfelt}
\affiliation{Department of Physics, Wesleyan University, Middletown, Connecticut 06459}
\author{Mei C. Zheng}
\affiliation{Department of Physics, Wesleyan University, Middletown, Connecticut 06459}
\affiliation{Department of Electrical Engineering, Princeton University, Princeton, New Jersey 08544}
\author{Ragnar Fleischmann}
\affiliation{Max Planck Institute for Dynamics and Self-organization (MPIDS),  37077 G\"{o}ttingen, Germany}
\author{Tsampikos Kottos}
\affiliation{Department of Physics, Wesleyan University, Middletown, Connecticut 06459}
\affiliation{Max Planck Institute for Dynamics and Self-organization (MPIDS),  37077 G\"{o}ttingen, Germany}
\begin{abstract}
We introduce a variant of the Banded Random Matrix ensemble and show, using detailed numerical analysis and theoretical 
arguments, that the phonon heat current in disordered quasi-one-dimensional lattices obeys a one-parameter scaling law. 
The resulting $\beta$-function indicates that an anomalous Fourier law is applicable in the diffusive regime, while 
in the localization regime the heat current decays exponentially with the sample size. Our approach opens a new way to 
investigate the effects of Anderson localization in heat conduction, based on the powerful ideas of scaling theory.
\end{abstract}
\maketitle
\section{Introduction} 
Anderson localization, i.e. the complete halt of propagation in disordered media due to wave interference effects, is an 
interdisciplinary field of research that addresses systems as diverse as classical, quantum and atomic-matter waves. This 
phenomenon was predicted fifty years ago in the framework of quantum (electronic) waves by Anderson \cite{A58} and its 
existence has been confirmed in recent years by experiments with classical \cite{WBLR97,CSG00,SGAM06,BZKKS09,HSPST08,C99,
LAPSMCS08,PPKSBNTL04,SBFS07} and matter waves \cite{A08,Ignuscio}.

Recently, localization phenomena due to randomness have attracted considerable interest in the context of heat conduction 
by phonons \cite{LLP03,D08,LXXZL12}. A central issue of these investigations is the determination of the dependence of the 
heat current $J$ on the system size $N$. It has been commonly believed that disorder scatters normal modes and induces a 
diffusive energy transport that leads to a normal heat conduction described by Fourier's law which states that $J\sim N^{-1}$. 
However, many recent studies \cite{D08,LXXZL12,LZH01,D01,DL08,LD05,RD08,KCRDLS10} suggest that in low dimensional disordered 
harmonic chains this may not always be true. Instead one finds that $J \sim N^{-\alpha}$, where $\alpha$ is usually different 
from one. Although this conclusion is generally accepted for one-dimensional systems, where theoretical methods of investigation
are available, the validity (or not) of Fourier law in higher dimensions is totally unclear since the majority of the available 
results are based on numerical simulations which are limited to small systems sizes.

In fact, recent experiments on heat conduction in nanotubes and graphene flakes have reported observations which indicate 
such anomalous behaviour with the system size \cite{COGMZ08,NGPB09,LRWZHL12}. Therefore, not only is it a fundamental demand 
for the development of statistical physics to understand normal and anomalous heat conduction in low dimensional systems, 
but it is also of great interest from the technological point of view, since the achievement of modern nano-fabrication 
technology allows one to access and utilize such structures with sizes in the range of a few nanometers up to few hundred nanometers.

In this Letter, we approach thermal transport in the presence of disorder from a different perspective, namely we develop a 
scaling theory for quasi-one-dimensional (1D) random lattices described by a modified Banded Random Matrix Ensemble (BRM). 
Random matrix models played a major role in understanding various properties of disordered quantum systems, including the structure 
and statistical properties of their eigenstates \cite{CMI90,BKS10} and eigenvalues \cite{CIM91}, the conductance \cite{CGIMZ94}, delay
times \cite{BBCK08}, etc. Here we introduce a BRM ensemble with bandwidth $2b+1$ that describes an array of coupled oscillators 
with long range ($b^\text{th}$-neighbor) random couplings, in the presence of on-site random pinning which is coupled at the left 
and right edges to a pair of Langevin heat reservoirs. We find that the averaged (rescaled) steady-state heat current ${\tilde J}_N
(\xi_{\infty})$ of the phononic excitations for an array of size $N$ obeys a one-parameter scaling, i.e.
\begin{equation}
\frac{\partial \ln {\tilde J}_N(\xi_{\infty})}{\partial \ln N} = \beta \left({\tilde J}_N(\xi_{\infty})\right),
\label{eq:scaling}
\end{equation}
where $\beta$ is a {\it universal} function of ${\tilde J}_N(\xi_{\infty})$ alone, and takes the following asymptotic forms
\begin{equation}
\label{sfunction}
\beta({\tilde J}_N)=\left\{
  \begin{array}{lcr}
    1.28 + 0.94 \ln {\tilde J}_N              & {\rm for} & {\tilde J}_N \ll 1 \\
    -\nu & {\rm for} & {\tilde J}_N \gg 1,
  \end{array}\right.
\end{equation}
with $\nu\approx 0.25$. The asymptotic (i.e. $N\rightarrow\infty$) participation number $\xi_{\infty}$ measures the degree of 
localization of the normal modes which dominate the transport. For any finite sample of size $N$ the number of these modes, $I$, scales as $I\sim 
N^{-\gamma}$ with $\gamma\approx 0.1$. The scaling exponent of the (actual) heat current $J_N\equiv N^{-\gamma}\cdot {\tilde J}_N 
\sim N^{-\alpha}$ is found to be $\alpha=\nu+\gamma\approx 0.35$, indicating a violation of the Fourier law. Eqs.(\ref{eq:scaling}-\ref{sfunction})
are confirmed in the following via detailed numerical simulations, supported by theoretical arguments.


\section{Banded Harmonic Chain Model}
We consider a thermally isolated quasi-one-dimensional harmonic oscillator chain with $b^\texttt{th}$-nearest neighbors coupling. 
It consists of $N$ equal masses $(m=1)$ described by the Hamiltonian
\begin{equation}
{\cal H}=\sum_{n=1}^{N} {\cal H}_n = \sum_{n=1}^{N} \left(\frac{p_n^2}{2}+\frac{\epsilon_n q_n^2}{2} + \frac{1}{4} \sum_{j=n-b}^{n+b} k_{nj} \left( q_n - q_j \right)^2
\right)\label{eq:H} 
\end{equation}
The corresponding equations of motion are $\dot{q}_n=\partial {\cal H}/\partial p_n, \; \dot{p}_n=-\partial {\cal H}/\partial q_n$,
where $q_n,p_n$ are respectively the individual oscillator displacements and momenta. The last term in Eq.~\ref{eq:H} is  the harmonic coupling
between the $n$-th mass and its $b$ neighbors on the left and right. The random spring constants $k_{nj}$ are chosen to be symmetric 
($k_{nj}=k_{jn}$) and uniformly distributed according to $k_{nj} \in \left[-\frac{W}{2}+1, \frac{W}{2}+1\right]$ if $0 < \abs{n-j} \leq b$, 
and $k_{nj}=0$ otherwise. $W$ is a coupling strength parameter that has to be smaller than 2 and is henceforth set to unity. 
The second term in the Hamiltonian is an on-site ``pinning'' potential with a spring constant $\epsilon_n$, random and uniformly
distributed in $\left[-\frac{W}{2}+1, \frac{W}{2}+1 \right]$. The offset in these random distributions ensure a positive-definite spectrum 
of the eigenfrequencies, i.e. bounded motion of the oscillators. The boundary conditions used are $q_0=q_{N+1}=0$.

\begin{figure}[hbt]
\centering
\includegraphics[keepaspectratio,width=\columnwidth]{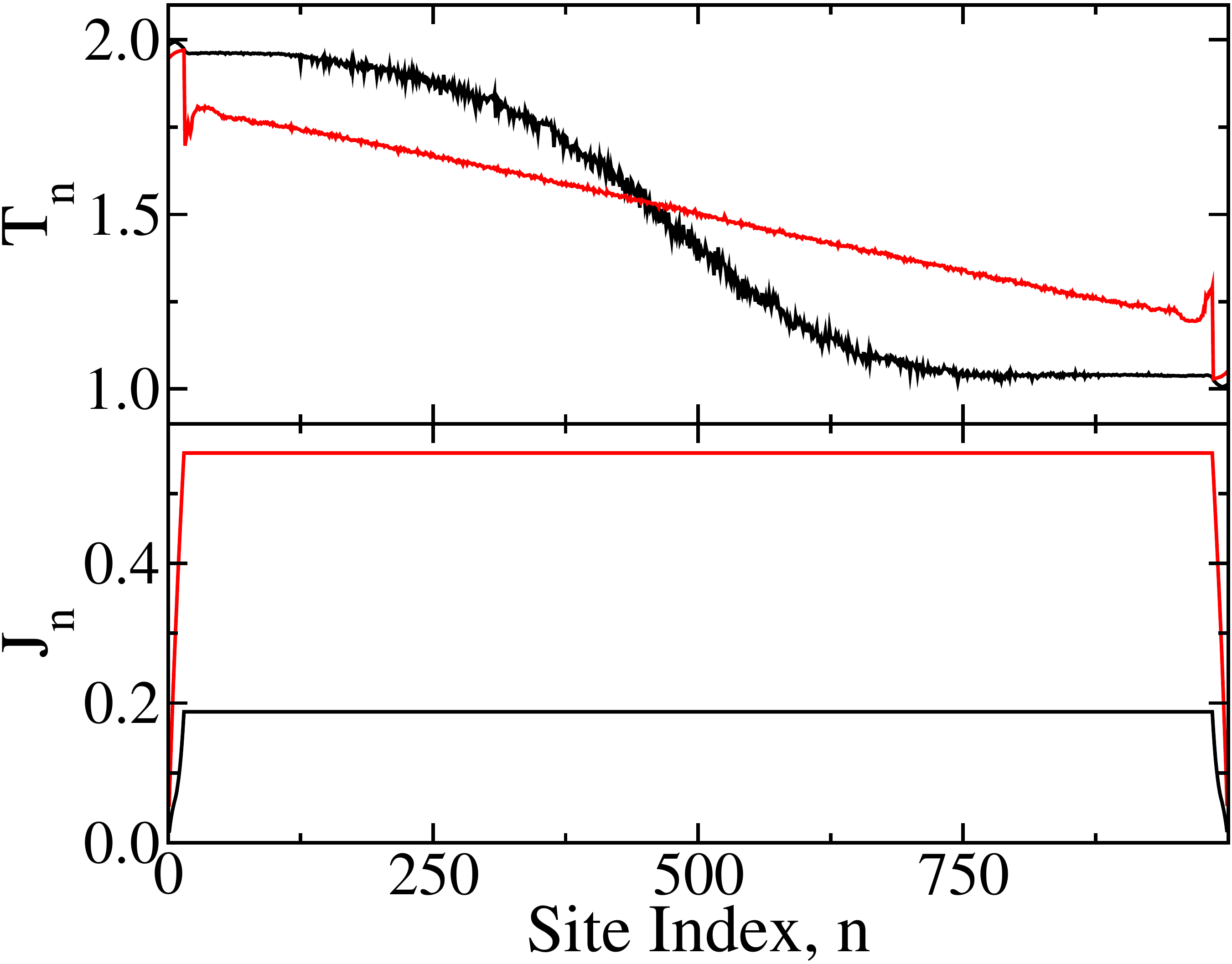} 
\caption{(Color online) Temperature (upper) and heat current (lower) profiles, for bandwidths of $b=6$ (black) and $b=30$ (red) in a system of size $N=1000$. 
The flatness of the heat current profile indicates a non-equilibrium steady state.}
\label{fig:tempflux}
\end{figure}
Next, we want to study the non-equilibrium steady states (NESS) of this chain driven by a pair of Langevin (Ornstein-Uhlenbeck) reservoirs 
set at temperature $T_L$ and $T_R$ respectively, and coupled to the first (last) $N_b$ masses with a constant coupling strength $\lambda$. 
In all numerical examples we will set $N_b=15$ and $\lambda=1$. The coupling to the bath is described by modifying the equation of motion 
for the momentum $ \dot{p}_n=-\partial {\cal H}/\partial q_n+\sum_{\tau=L,R} \left(-\lambda p_n + \sqrt{2\lambda T_\tau} \zeta_n\right) 
\theta_n^\tau$, where $\theta_n^L=\lbrace 1 \text{ if } n \leq N_b;  0 \text{ otherwise} \rbrace$, $\theta_n^R=\lbrace 1 \text{ if } n 
\geq N-N_b; 0 \text{ otherwise}\rbrace$, and $\zeta_n(t)$ is delta-correlated white noise $\avg{\zeta_n(t) \zeta_{n'}(t')} = \delta_{nn'} \delta(t-t')$.

The two thermal quantities, local temperature and the heat current, can be expressed in terms of elements of 
the covariance matrix $\Cb(t) = \left\langle \vec{x}(t) \otimes \vec{x}(t)\right\rangle$, where $\vec{x} = (q_1,\ldots q_N, p_1, \ldots p_N)^T$ 
is the state vector. Using stochastic Ito calculus \cite{LLP03,ZEKFGP11} for the system of Eq.(\ref{eq:H}) we find
\begin{equation}
\frac{d\Cb}{dt} = \Zb \Cb + \Cb \Zb^{T} + \Yb, \label{eq:cov}
\end{equation}
with the $2N \times 2N$ matrices 
\begin{equation}
\Zb = \left( \begin{array}{cc} \bz & \bo \\ \Kb & \bo \end{array} \right) - \sum_{\tau=L,R} \Yb^\tau \text{and } 
\Yb = \sum_\tau T_\tau \Yb^\tau,\label{eq:Z} 
\end{equation}
where $\Yb^\tau = \lambda \sum_{n=1}^N \theta_n^\tau \Pb_{N+n}$ and $\Pb_j = \vec{e}_j \otimes \vec{e}_j$ is a diagonal rank 1 projector for 
basis vectors $\left(\vec{e_j}\right)_n = \delta_{nj}$. The banded $N\times N$-matrix $\Kb$ with bandwidth $2b+1$ encodes all the interactions 
within the harmonic lattice, as described by Eq.(\ref{eq:H}) 
\begin{equation*}
K_{nm} = \begin{cases}
           k_{nm} & \quad n \ne m, \\
           -\epsilon_n -\sum_{j} k_{nj} & \quad n = m 
          \end{cases}
\end{equation*}
The NESS covariance matrix, $\Cb^\infty$, can be obtained by setting the left hand side of Eq.(\ref{eq:cov}) to zero, resulting in the 
\emph{Sylvester equation} $\Zb \Cb^\infty + \Cb^\infty \Zb^{T} = -\Yb$.

The local temperature is simply given by $T_n = \left<p_n^2\right> = \Css_{n+N, n+N}$. To find the expression for the heat current $J$ we use 
the continuity relation $\partial_t \left<{\cal H}_n\right>+J_n-J_{n-1}=0$, where $\left< {\cal H}_n\right>$ is the thermal fluctuation average of
\begin{widetext}
\begin{equation}
\frac{\partial {\cal H}_n}{\partial t} = \frac{1}{2} \sum_{j>0}^b k_{n+j,n} \left( q_{n+j}-q_n \right) \left( p_n+p_{n+j} \right) - k_{n-j,n} \left( q_n-q_{n-j} \right) \left(
p_n+p_{n-j} \right) \label{eq:jn},
\end{equation} 
ensuring that at any given cross section of the chain, all connections are included (see supplemental material for details). In terms of 
the covariance matrix this yields the expression
\begin{equation}
J_n = \frac{1}{2} \sum_{i=n-b+1}^n \sum_{j=n+1}^{i+b} k_{i-n+b,j-n} \left( \Css_{i,i+N}  - \Css_{j,i+N} + \Css_{i,j+N} -  \Css_{j,j+N} \right). \label{eq:current}
\end{equation}
\end{widetext}
In the central section of the chain ($N_b<n<N-N_b$), which is not directly coupled to the bath, the NESS heat current has to be independent 
of $n$ due to continuity, i.e.  $J=J_n$. In our proceeding numerics, the bath temperatures are fixed \cite{footnote1} at $T_L=2, \; T_R=1$. 
Additionally, all thermal calculations are averaged over $10^2$ realizations of disorder. An example of the local profiles is displayed in 
Fig.\ref{fig:tempflux} -- in particular, the flat profile seen in $J_n$ confirms that continuity is fulfilled.  

\section{Localization Properties} 
\begin{figure}[h]
\centering
\includegraphics[keepaspectratio,width=\columnwidth]{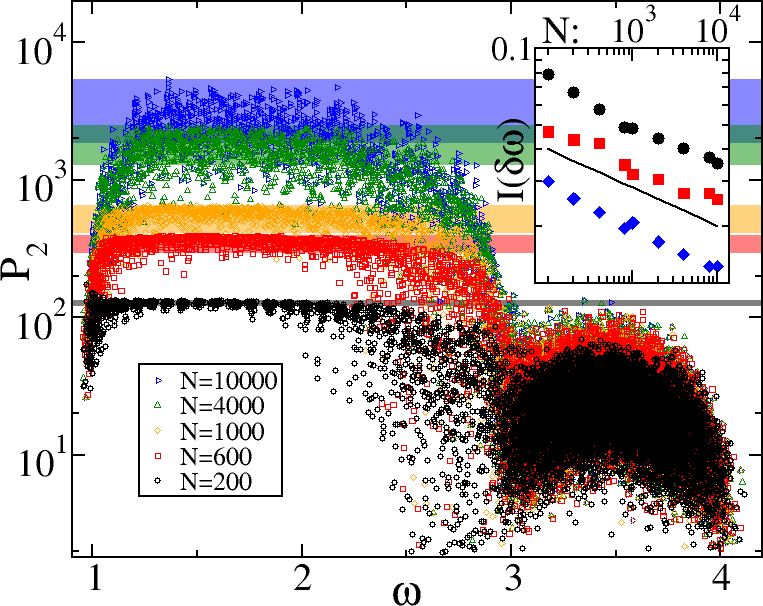} 
\caption{(Color online) Participation number (PN) versus frequency for $b=5$. Various finite system sizes are delineated by colors. Colored stripes indicate 
the groups of the most extended modes, for which $P_2 \ge 0.6 P_2^{\rm max}$. A normalized count of selected states yield a scaling $I\sim N^{-\gamma}$ 
for the integrated density of states (IDoS), as shown in the inset, for three representative bandwidths: $b=5$ (black $\bullet$), $b=10$ (red 
$\blacksquare$), and $b=12$ (blue $\blacklozenge$). The best fit indicates $\gamma\approx 0.1\pm 0.05$ (black solid line).} 
\label{fig:frq_v_IPR}
\end{figure}

We consider the isolated case ($\lambda=0$). Substituting $q_n^\nu(t) = A_{n,\nu} \exp\left( i \omega_\nu t \right)$ results in an eigenvalue 
problem $-\omega_\nu^2 {\vec A}_\nu = \Kb {\vec A}_\nu$. Again, note that the choice of the random distributions in the banded random matrix $\Kb$ 
ensures positive definite eigenvalues, $\omega_\nu^2 \ge 0$. The extent of the modes is often characterized by their participation number (PN)
\begin{equation}
P_2 = \frac{ \left( \sum_n \abs{ A_{n,\nu} }^2 \right)^2 }{ \sum_n \abs{ A_{n,\nu} }^4 }. \label{eq:P2}
\end{equation}
In Fig.\ref{fig:frq_v_IPR}, the PN's  are plotted versus the eigenfrequencies for different finite lattice sizes and a fixed bandwidth of $b=5$. 
All states are localized ($P_2<N$), yet two windows are observed: a window of highly localized states for higher frequencies and a window of 
states with larger PN for lower frequencies.

\begin{figure}[h]
\centering
\includegraphics[keepaspectratio,width=\columnwidth]{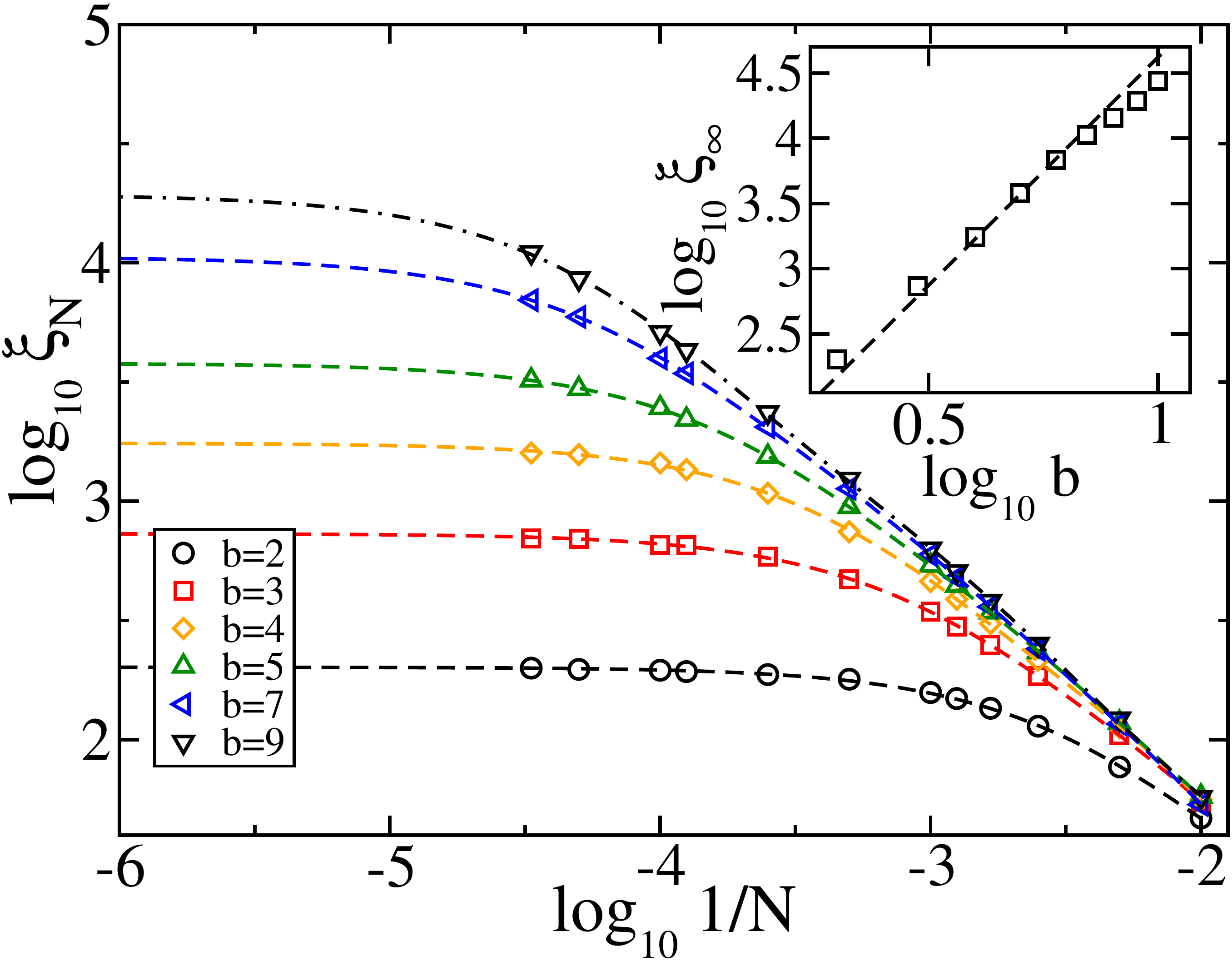}
\caption{(Color online) \textit{Main:} The disordered and spectral (over $\delta \omega$) averaged PN $\xi_N$ against $1/N$, for various $b$-values. 
Saturation of $\xi_N$ as $N\rightarrow\infty$ is observed for moderate bandwidths. For larger values of $b$ a fitting of the data to a 
rational function of fifth order polynomials (dashed/dot-dashed lines) allows to extract the limit $\xi_\infty$. \textit{Inset:} Parametric 
dependence of $\xi_\infty$ on $b$. The dashed line is the power relation $b^{3.5}$.}
\label{fig:asymptotic}
\end{figure}

We define the spectral window $\delta\omega=\omega_{\rm max}-\omega_{\rm min}$, whereby the modes with the larger PN are supported by the 
condition $P_2(\omega) \ge 0.6 P_2^{\rm max}$. This allows us to find a scaling behavior for the integrated density of states (IDoS) of these 
modes 
\begin{equation}
I\left(\delta\omega\right) = \int_{\omega_\text{min}}^{\omega_\text{max}} \rho(\omega) d\omega \sim N^{-\gamma}, \label{eq:IDoS}
\end{equation}
Our results for various values of bandwidth $b$ are shown in the inset of Fig.\ref{fig:frq_v_IPR}. The solid line indicates the best fit with 
$\gamma =0.1\pm 0.05$. The averaged (over the spectral window $\delta \omega$ and over disorder realizations) PN is $\xi_N=\avg{P_2}_{\delta\omega}$, 
reported in Fig.\ref{fig:asymptotic} for various $b$-values versus $1/N$. Typically, more than $10^{4}$ eigenvectors were used for the averaging. 
We find that $\xi_{N\rightarrow\infty}(b)$ shows a convergence towards a finite value $\xi_{\infty}(b)$. For moderate $b$-values this asymptotic 
PN is reached, while for larger bandwidths it can be extrapolated from the quotient of two fifth order polynomials fitted to the data (dashed/dotted
-dashed lines in the figure). Inspired by previous studies on the localization properties of BRM's \cite{CMI90,KPIR96,BKS10}, we speculate that the
asymptotic PN will scale as $\xi_\infty \sim b^\eta$. Our expectation is nicely confirmed by the numerical data, reported in the inset of 
Fig.~\ref{fig:asymptotic}. The best fit indicates that $\eta \approx 3.5$, and thus $\xi_\infty(b) \sim b^{3.5}$.


\section{Scaling Theory}
Equipped with knowledge of the localization properties of the normal modes of our system in Eq.(\ref{eq:H}), we now turn to the study 
of the steady-state heat current of Eq.(\ref{eq:current}). This is formally expressed as $J = \int_{0}^{\infty} \rho(\omega) \tau(\omega) 
d\omega\approx \int_{\omega_{\rm min}}^{\omega_{\rm max}} \rho(\omega) \tau(\omega) d\omega,$ where $\rho(\omega)$ is the density of 
states and $\tau(\omega)$ is the frequency dependent transmittance. Since heat is transported significantly only by the modes with the
larger PN, we confine the integration range within the spectral window $\delta\omega$. These modes have similar localization -- and 
therefore transport -- properties, as shown previously. Therefore, we approximate $\tau(\omega)$ by its average value over this window 
$\avg{\tau}_{\delta\omega}$. The remaining integral is then just the IDoS of Eq.(\ref{eq:IDoS}). We can now use knowledge for the 
transmittance of harmonic chains to deduce a scaling relation for the rescaled heat current ${\tilde J}\equiv J/N^{-\gamma} \propto 
\avg{\tau}_{\delta\omega}$.

\begin{figure}[h]
\centering
\includegraphics[keepaspectratio, width=\columnwidth]{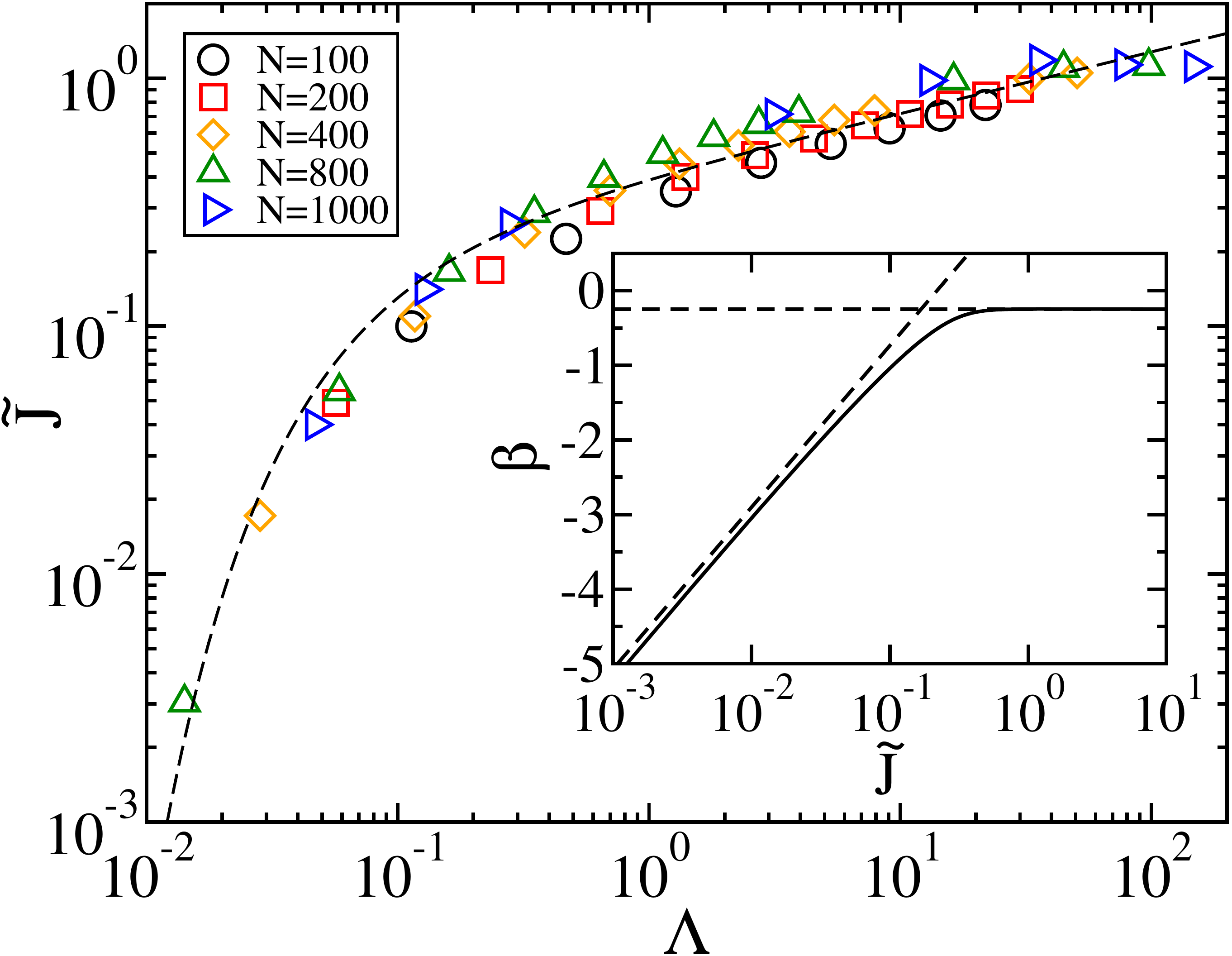}
\caption{(Color online) \textit{Main:} Rescaled heat current ${\tilde J}$ vs. $\Lambda = \xi_{\infty}/N$ where $\xi_{\infty}\sim b^{3.5}$ as was 
found from Fig.\ref{fig:asymptotic}. The dashed line is a fit to the analytical curve of Eq.(\ref{eq:analytic_scaling}). \textit{Inset:} 
The resulting $\beta$-function of Eq.(\ref{eq:scaling}). The dashed lines correspond to the asymptotic limits of $\beta = 1.28 + 0.94 
\cdot \ln {\tilde J}$ for $N \rightarrow \infty$ and $\beta = -\nu$ for $N\rightarrow 0$.}
\label{fig:scaled}
\end{figure}

Specifically, the transport theory of disordered media predicts that the average transmittance $\avg{\tau}_{\delta\omega}$ of a 
disordered sample of length $N$ which is characterized by a localization length $\xi_{\infty}$ follows a one-parameter scaling 
$\avg{\tau}_{\delta\omega}=f_T(\Lambda)$, where the one parameter is $\Lambda\equiv \xi_{\infty}\ / N$. It is natural to then speculate 
that the same scaling relation will apply for the rescaled heat current ${\tilde J}$. In the main panel of Fig.\ref{fig:scaled} we 
show our numerics of ${\tilde J}$ plotted against $\Lambda$, for a number of different bandwidths and system sizes ($b \in [2,35], N 
\in [10^2, 10^3]$). In this approach, we have used $\xi_\infty$ as a scaling parameter, which allows us to collapse all 
data associated with various $b$-values to one scaling curve. By visual inspection~\cite{footnote2}, we find that the scaling parameter $\xi_\infty \sim b^{3.5 \pm 0.2}$, 
which confirms the previous independent scaling analysis from the participation numbers (see Fig.\ref{fig:asymptotic} and its discussion). 
The obvious data collapse confirms the conjecture that ${\tilde J}_N(\xi_{\infty})$ is a function of $\Lambda$ only i.e. 
${\tilde J}_N(\xi_{\infty}) =f_{\tilde J}(\Lambda)$. 

Next we want to determine the analytical form of the scaling function $f_{\tilde J}(\Lambda)$. We have found that in the limit 
of the localized regime ($\Lambda \ll 1$) this dependence has the form ${\tilde J}\sim e^{-c_0 / \Lambda}$, in agreement with 
previous theoretical results for pinned harmonic chains with only nearest neighbor coupling and mass disorder \cite{DL08}. In the 
other limit of $\Lambda \gg 1$, the heat transport is diffusive. Assuming validity of the Fourier law, we may expect a scaling of 
the type ${\tilde J}\sim 1/N^{1-\gamma}$ -- however, recent investigations \cite{lepri_stochastic_2009, KCRDLS10} found an anomalous 
behavior of the heat current, which results in the scaling ${\tilde J}\sim 1/N^{\alpha-\gamma}$. We therefore speculate that in the 
diffusive domain of $\Lambda\gg1$, the rescaled steady state heat current will follow the relation ${\tilde J}(\Lambda) \sim c_1 
\Lambda^{\nu}$. A possible interpolating law valid in all regimes (including the crossover region) is
\begin{equation}
{\tilde J}(\Lambda) =  (c_2 + c_1 \Lambda^{\nu} )\exp(-c_0 / \Lambda). \label{eq:analytic_scaling}
\end{equation}
Comparison with numerical data in the two limits ($\Lambda\gg 1$, $\Lambda\ll 1$) yields $\nu\approx 0.25$, $c_1\approx 0.4$ and 
$c_0\approx 0.06$. Adjusting the last parameter $c_2$ to fit numerics in the intermediate region yields $c_2\approx 0.012$. The 
resulting analytical formula nicely fits the numerical results in all regions, and therefore provides a compact summary of our 
empirical data (see dashed line in Fig.\ref{fig:scaled}). We stress that the limiting value of $J(\Lambda\gg 1)\propto
1/N^{\gamma+\nu}$ leads to an anomalous heat exponent $\alpha\approx 0.35\pm 0.05$. It should be noted that this value is less 
than what has typically been seen in other - nonlinear and disordered - chain models, which show values from $\approx 0.4-0.7$  
\cite{ivanchenko_disorder_2011,mai2007,D01,lepri_stochastic_2009,RD08,KCRDLS10}.

Eq.(\ref{eq:analytic_scaling}) can be rewritten in the form of Eq.(\ref{eq:scaling}). This is the main result of the present Letter, 
as it allows postulating the existence of a $\beta$-function for the ${\tilde J}_N$ of \textit{generic} quasi-$1D$ disordered systems. 
The resulting $\beta$-function is plotted in the inset of Fig.\ref{fig:scaled}. Its asymptotes are seen to follow $\beta = 1.28 + 0.94 
\cdot\ln {\tilde J}$ for $N\rightarrow\infty$, and $\beta = -\nu$ for $N\rightarrow 0$.

\section{Conclusions} 
We presented a one-parameter scaling theory for the steady-state heat current of quasi-one dimensional disordered harmonic systems 
with substrate pinnings described by a variant Banded Random Matrix ensemble. Via numerical analysis and theoretical considerations, 
we have established Eq.(\ref{eq:scaling}), which allows us to conclude that changing disorder strength (or coupling range) and system 
size in the way described by Eq.(\ref{eq:analytic_scaling}), would not change the renormalized (average) heat current. The one-parameter
scaling theory presented here is a powerful approach in the quest of understanding thermal transport, and validity of the Fourier law, 
in disordered media. Of further interest will be to investigate higher moments of the NESS heat current and also to establish a scaling theory 
for the thermal profile, as a function of the scaling parameter $\Lambda$. Although the focus of this Letter was on harmonic quasi-1D disordered 
phononic transport, our approach can be used to study high-dimensional pinned harmonic systems, and to better understand the effects of phonon-phonon 
interactions \cite{BKS10} in thermal transport.

\acknowledgments{The authors wish to thank T. Prosen and S. Flach for useful discussions. This research was supported by an AFOSR No.
FA 9550-10-1-0433 grant, by a NSF ECCS-1128571 grant, and by the DFG Forschergruppe 760.}

\begin{thebibliography}{99}

\bibitem{A58} P. Anderson, Phys. Rev. {\bf 109}, 1492 (1958).

\bibitem{WBLR97} D.S. Wiersma et al., Nature {\bf 390}, 671 (1997).

\bibitem{CSG00} A.A. Chabanov, M. Stoytchev, \& A.Z. Genack, Nature {\bf 404}, 850 (2000).

\bibitem{SGAM06} M. St\"orzer et al., Phys. Rev. Lett. {\bf 96}, 063904 (2006).

\bibitem{BZKKS09} J.D. Bodyfelt et al., Phys. Rev. Lett. {\bf 102}, 253901 (2009).

\bibitem{HSPST08} H. Hu et al., Nature {\bf 4}, 945 (2008).

\bibitem{C99} H. Cao et. al., Phys. Rev. Lett. {\bf 82}, 2278 (1999); H. Cao, Waves in Random Media {\bf 13}, R1 (2003).

\bibitem{LAPSMCS08} Y. Lahini et al., Phys. Rev. Lett {\bf 100}, 013906 (2008).

\bibitem{PPKSBNTL04} T. Pertsch et al., Phys. Rev. Lett. {\bf 93}, 053901 (2004).

\bibitem{SBFS07} T. Schwartz et al., Nature {\bf 446}, 52 (2007).

\bibitem{A08} A. Aspect et al., Nature {\bf 453}, 891 (2008).

\bibitem{Ignuscio} G. Roati et al., Nature {\bf 453}, 895 (2008).

\bibitem{LLP03} S. Lepri, R. Livi, \& A. Politi, Phys. Rep. {\bf 377}, 1 (2003).

\bibitem{D08} A. Dhar, Adv. Phys. {\bf 57}, 457 (2008).

\bibitem{LXXZL12} S. Liu et al., arXiv:1205.3065v2 [cond-mat.stat-mech] (2012).

\bibitem{DL08} A. Dhar \& J.L. Lebowitz, Phys. Rev. Lett. {\bf 100}, 134301 (2008).

\bibitem{LD05} L. W. Lee \& A. Dhar, Phys. Rev. Lett. {\bf 95}, 094302 (2005).

\bibitem{RD08} D. Roy \& A. Dhar, Phys. Rev. E {\bf 78}, 051112 (2008).

\bibitem{LZH01} B. Li, H. Zhao, \& B. Hu, Phys. Rev. Lett. {\bf 86}, 63 (2001).

\bibitem{KCRDLS10}A. Kundu et al., Europhys. Lett. {\bf 90}, 40001 (2010);
A. Chaudhuri et al., Phys. Rev. B {\bf 81}, 064301 (2010).

\bibitem{D01} A. Dhar, Phys. Rev. Lett.  {\bf 86}, 5882 (2001).

\bibitem{COGMZ08} C.W. Chang et al., Phys. Rev. Lett, {\bf 101}, 075903 (2008);
G. Zhang \& B. Li, NanoScale {\bf 2}, 1058 (2010).

\bibitem{NGPB09} D. L. Nika et al., Appl. Phys. Lett. {\bf 94}, 203103 (2009).

\bibitem{LRWZHL12} N. Li et al., Rev. Mod. Phys. \textbf{84}, 1045 (2012).

\bibitem{CMI90} G. Casati, L. Molinari, \& F. Izrailev, Phys. Rev. Lett. {\bf 64}, 1851 (1990);
Y. V. Fyodorov \& A. D. Mirlin, Phys. Rev. Lett. {\bf 69}, 1093 (1992); Phys. Rev. Lett. {\bf 67}, 2405 (1991);  
Int. J. Mod. Phys. {\bf 8}, 3795 (1994).

\bibitem{KPIR96} T. Kottos et al., Phys. Rev. E \textbf{53}, R5553 (1996); T. Kottos, A. Politi, \& F.M. Izrailev,
J. Phys.: Cond. Matt. \textbf{10}, 5965 (1998).

\bibitem{BKS10} J.D. Bodyfelt, T. Kottos, \& B. Shapiro, Phys. Rev. Lett. {\bf 104}, 164102 (2010).

\bibitem{CIM91} G. Casati, F. Izrailev \& L. Molinari, J. Phys. A {\bf 24}, 4755 (1991).

\bibitem{CGIMZ94} G. Casati et al., Phys. Rev. Lett. {\bf 72}, 2697 (1994); T. Kottos, F. Izrailev, \& A. Politi, Physica D  {\bf 131}, 155 (1999).

\bibitem{BBCK08} J.D. Bodyfelt et al., Phys. Rev. B {\bf 77}, 045103 (2008); A. Ossipov \& Y. V. Fyodorov, Phys. Rev. B {\bf 71}, 125133 (2005).

\bibitem{ZEKFGP11} M.C. Zheng et al., Phys. Rev. E {\bf 84}, 021119 (2011).

\bibitem{Evangelou90} S.N. Evangelou \& E.N. Economou, Phys. Lett. A {\bf 151}, 345–348 (1990).

\bibitem{izrailev_periodic_1996} F.M. Izrailev, L. Molinari, \& K. \.{Z}yczkowski, J. de Physique I {\bf 6}, 455–468 (1996).

\bibitem{lepri_stochastic_2009} S. Lepri, C. Mejia-Monasterio, \& A. Politi, J. Phys. A: Math. and Theor. {\bf 42}, 025001 (2009).

\bibitem{ivanchenko_disorder_2011} M.V. Ivanchenko \& S. Flach, Europhys. Lett. \textbf{94}, 46004 (2011).

\bibitem{mai2007}  T. Mai, A. Dhar, \& O. Narayan, Phys. Rev. Lett. {\bf 98}, 184301 (2007).

\bibitem{footnote1} We checked that our results do not sensitively depend on this particular choice of temperatures, by repeating the 
calculations with different pairs of bath temperatures on the same order of magnitude.

\bibitem{footnote2} In most cases, error bars are on the order of the symbol size, apart from $N=800,1000$, where the error bars are slightly 
larger (approximately twice the size). These data sets are the most numerically costly, and therefore were performed with less realizations.  

\end{thebibliography}

\end{document}


\section*{Supplemental Material}
\subsection*{Expression for the Heat Current in Long-Range Connected Oscillator Chains}
\begin{figure}[h]
\centering
\includegraphics[keepaspectratio,width=0.9\textwidth]{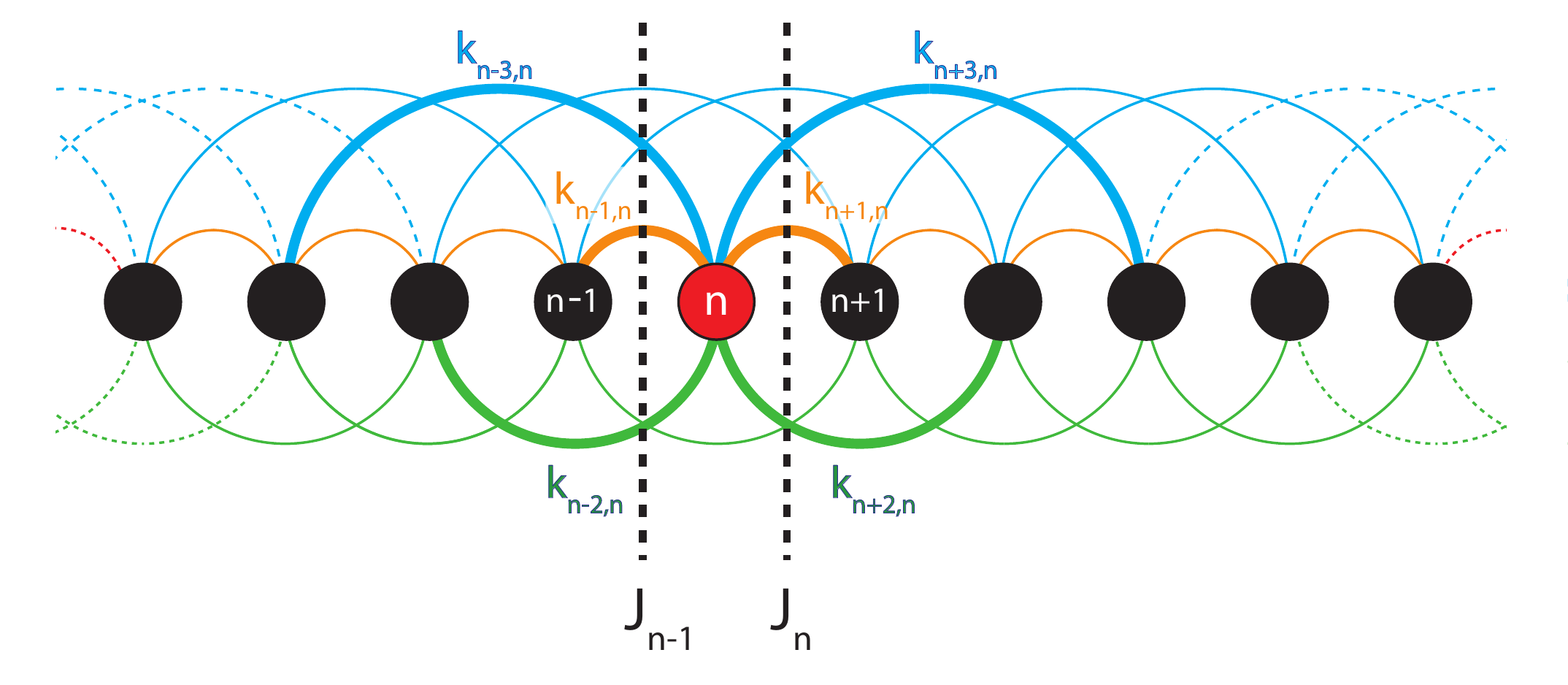} 
\caption{Illustration of the contributions to the thermal current for a banded chain model with b=3, i.e.\ each oscillator is coupled to the three neighbors on each side.}
\label{fig:currentdef}
\end{figure}

The usual way to find the correct expression for the heat current in oscillator chains is to start with the continuity relation
\begin{equation}
\frac{\partial \left<{\cal H}_n\right>}{\partial t}+J_n-J_{n-1}=0
\end{equation}
that connects the local energy density change with the net heat current. The brackets $\left<\cdot\right>$ indicate the average over the bath stochasticity.
By examining the expression for the change in local energy 
\begin{equation}
\frac{\partial {\cal H}_n}{\partial t} = \frac{1}{2} \sum_{j>0}^b k_{n+j,n} \left( q_{n+j}-q_n \right) \left( p_n+p_{n+j} \right) - k_{n-j,n} \left( q_n-q_{n-j} \right) \left(
p_n+p_{n-j} \right) \label{eq:dhdt}
\end{equation} 
one can identify the terms contributing to $J_n$ and $J_{n-1}$, respectively.
In our chain with longer than nearest neighbor coupling, however, this procedure misses terms in the heat current, as illustrated in Fig.~\ref{fig:currentdef}: Let us examine the
current $J_n$ at the position indicated by the right dashed vertical line. Apart from the couplings from site $n$ to its neighbors on the right (thick lines) there are other
couplings penetrating the vertical line. Those connections transport heat through this section as well and thus have to be accounted for in the expression for $J_n$. From
Eq.~\ref{eq:dhdt} we can see, that every coupling contributes a term $- \left<k_{i,j} (q_i-q_j)(p_i+p_j)\right>$ to the current. We thus find
\begin{equation}
J_n=-\sum_{\substack{i>n,j\le n \\ i-j\le b}} k_{i,j} \left<(q_i-q_j)(p_i+p_j)\right>,
\end{equation}     
that can easily be expressed in terms of entries of the covariance matrix as given in the article. From Fig.~\ref{fig:currentdef} we see, that the thin lines penetrating the right
dashed vertical line penetrate the left one as well, therefore they drop out of the expression $J_n-J_{n-1}$ and do not occur in the continuity equation.  
